\documentclass[a4paper]{article}
\usepackage{amsmath}

\renewcommand{\r}[1]{(\ref{#1})}

\newcommand{\dt}{\! \cdot \!}
\newcommand{\nn}{\nonumber}
\newcommand{\bx}{\mbox{\boldmath $x$}}
\newcommand{\bp}{\mbox{\boldmath $p$}}
\newcommand{\bk}{\mbox{\boldmath $k$}}
\newcommand{\bq}{\mbox{\boldmath $q$}}
\newcommand{\grad}{\nabla}
\newcommand{\bgrad}{\mbox{\boldmath $\grad$}}
\newcommand{\clm}{{\mathcal{M}}}
\newcommand{\eps}{\epsilon}
\newcommand{\lam}{\lambda}
\newcommand{\delsl}{\not\!\partial}
\newcommand{\psl}{\not\!p}
\newcommand{\ksl}{\not\!k}

\begin{document}

\begin{center}

{\bf\Large Perturbation Theory Calculation of the Black \\ 
Hole Elastic Scattering Cross Section } 

\vspace{0.4cm}

Chris Doran\footnote{e-mail: \texttt{c.doran@mrao.cam.ac.uk}, 
\texttt{http://www.mrao.cam.ac.uk/$\sim$cjld1/}}
and Anthony Lasenby\footnote{e-mail: \texttt{a.n.lasenby@mrao.cam.ac.uk}}

\vspace{0.4cm}

Astrophysics Group, Cavendish Laboratory, Madingley Road, \\
Cambridge CB3 0HE, UK.

\vspace{0.4cm}

\begin{abstract}
The differential cross section for scattering of a Dirac particle in a
black hole background is found.  The result is the gravitational
analog of the Mott formula for scattering in a Coulomb background.
The equivalence principle is neatly embodied in the cross section,
which depends only on the incident velocity, and not the particle
mass.  The low angle limit agrees with classical calculations based on
the geodesic equation.  The calculation employs a well-defined
iterative scheme which can be extended to higher orders.  Repeating
the calculation in different gauges shows that our result for the
cross section is gauge-invariant and highlights the issues involved in
setting up a sensible iterative scheme.
\end{abstract}

\vspace{0.4cm}


\end{center}

\section{Introduction}

Scattering of a charged fermion in a background Coulomb field is a
widely studied process which can be analysed perturbatively in quantum
field theory.  To lowest order the scattering process is summarised in the
Mott formula for the unpolarised differential scattering cross
section~\cite{itz-quant},
\begin{equation}
\left. \frac{d\sigma}{d\Omega} \right|_{\mbox{Mott}} = \frac{Z^2
\alpha^2}{4 \bp^2 v^2 \sin^4(\theta/2)} \bigl(1-v^2 \sin^2(\theta/2)
\bigr). 
\end{equation}
Here $\bp$ is the momentum, $v=|\bp|/E$, $\alpha$ is the fine
structure constant and the source has charge $Ze$.  Curiously, there
have only been sporadic attempts to repeat the analysis leading to the
Mott formula for the case of a black hole~\cite{col73,san78}.  The
problem of scattering by a black hole has certainly been tackled by
many authors (see for example the books by Futterman, Handler and
Matzner~\cite{futt-sct} and Chandrasekhar~\cite{cha83}, or the recent
article by Andersson and Jensen~\cite{and00}).  But few authors tackle
the problem in perturbation theory and there is a notable reluctance
to analyse the fermion case, with most work carried out for the case
of a massless scalar field.

Unlike scattering in a Coulomb field, black hole scattering is
complicated by the additional effects of absorption and emission.
Absorption is due to the singularity in the gravitational field and
manifests itself as a lack of Hermiticity in the fermion wave
equation~\cite{san77,DGL98-grav,D00-kerr}.  Emission is due to the
Hawking radiation and cannot effectively be treated without some of
the apparatus of quantum field theory~\cite{bir-quant}.  Despite these
complicating factors, we show here that an iterative scheme can be set
up, based on the Dirac equation, which produces a formula for the
lowest-order scattering cross section with little difficulty.

In this paper we concentrate on the scattering cross section for a
fermion in the background field of a spherically-symmetric black hole.
We find that the unpolarised differential scattering cross section is
given by
\begin{equation}
\frac{d\sigma}{d\Omega} = \frac{(GM)^2}{4 v^4 \sin^4(\theta/2)} \bigl(
1 + 2v^2 - 3 v^2 \sin^2(\theta/2) + v^4 - v^4 \sin^2(\theta/2) \bigr).
\label{sctint}
\end{equation}
We calculate this result in two quite different gauges, which require
going to different orders in perturbation theory.  The first
calculation utilises a series of gauge choices for the gravitational
fields which convert the Dirac equation into a simple, Hamiltonian
form~\cite{DGL98-grav,D00-kerr}.  This can be analysed
straightforwardly through a perturbation expansion.  Unlike approaches
based on the Schwarzschild coordinate system, the gauge used here does
not contain a singularity at the horizon and extends right up to the
origin.  This means that each step in the perturbation series can be
treated exactly,  avoiding the problems encountered by Collins
\textit{et al.}~\cite{col73} who attempted a Born approximation
scheme based on Schwarzschild coordinates.  The result they obtained
was physically unreasonable, though their explanation of why this
occurred (as the result of wavepacket dispersion) seems incorrect in
light of the present calculation.

The choice of gauge employed in our first calculation leads to an
unusual form of the vertex factor in momentum space, which vanishes
when both the incoming and outgoing fermions are on-shell.  It follows
that the first-order contribution to the scattering cross section in
this gauge is identically zero.  Since the vertex factor turns out to
go as the square root of the black hole mass, the fact that the
process is second-order does make sense.  The integrals involved in
the second-order calculation are all finite and do not require any
form of regularisation procedure.  The result of the
calculation~\r{sctint} is the gravitational analog of the Mott
scattering formula.  For low energies the cross section reduces to the
familiar Rutherford formula.  As the energy is increased, relativistic
corrections become more significant.  The formula neatly embodies the
equivalence principle, in that the cross-section depends only on the
particle's velocity, and not its mass.  In the low-angle limit our
formula agrees with earlier results for the classical cross-section
based on the geodesic equation~\cite{col73}.

The success of the calculation suggests a more general scheme for
tackling the scattering problem, and this is explored in the second
half of this paper.  We calculate the cross section in a different
gauge, using a first-order scheme, and confirm that the same result is
obtained.  This verifies that the cross-section formula is
gauge-invariant.  The second calculation also highlights the
conceptual difficultly of carrying through a full programme based on
perturbation theory, as choice of gauge at the start of the
calculation can dictate features as important as the order of Feynman
diagrams one needs to consider.  This appears to be the correct
explanation for the failure of the Collins \textit{et
al.}~\cite{col73} calculation.  Their scheme was based on the
Schwarzschild gauge, and they only went to first order in perturbation
theory.  But to be consistent we should include terms from the
Schwarzschild metric to first order in $M$, which then recovers the
correct result.

Throughout we employ units with $c=\hbar=1$, while factors of $G$
are stated explicitly.  Where appropriate, factors of $c$ and $\hbar$
are also included.  The Minkowski spacetime metric has signature
$(1,-1,-1,-1)$.

\section{The Dirac Equation}

Our starting point is the Schwarzschild line element in its standard form
\begin{equation}
ds^2 = \left( 1-\frac{2GM}{r} \right) d\bar{t}^2 - \left(
1-\frac{2GM}{r} \right)^{-1} dr^2 -
r^2( d\theta^2 + \sin^2\!\theta \, d\phi^2),
\label{swzmt}
\end{equation}
where $\bar{t}$ is the proper time measured by stationary observers,
and $r$, $\theta$ and $\phi$ have their usual meaning.  We first
transform to a new time coordinate $t$, corresponding to the proper
time for particles infalling radially from rest at infinity.  This
coordinate has
\begin{equation}
dt = d\bar{t} + \frac{(2GMr)^{1/2}}{r-2GM} dr.
\end{equation}
In terms of this new coordinate the line element~\r{swzmt} becomes
\begin{equation}
ds^2 = d t^2 - \Bigl(dr + \Bigl(\frac{2GM}{r} \Bigr)^{1/2} dt \Bigr)^2 -
r^2( d\theta^2 + \sin^2\!\theta \, d\phi^2).
\label{swzmt2}
\end{equation}
The coordinate system is valid for $0\leq r<\infty$.  We will not
consider the effects of analytic continuation through the introduction
of Kruskal-Szekeres coordinates.  See~\cite{DGL98-grav, D00-kerr} for
a more detailed discussion of the form of the metric employed here.

In order to write the Dirac equation in its simplest form it is useful
to revert to Cartesian coordinates by writing
\begin{align}
x &= r \sin \! \theta \cos \! \phi \nn \\
y &= r \sin \! \theta \sin \! \phi \nn \\
z &= r \cos \! \theta.
\end{align}
With the coordinates written $x^\mu=(t,x,y,z)$, $\mu=0\ldots 3$, the
line element~\r{swzmt2} can be written as
\begin{equation}
ds^2 = \eta_{\mu\nu} dx^\mu dx^\nu - \frac{2GM}{r}dt^2 - \frac{2}{r} \left(
\frac{2GM}{r} \right)^{1/2} x^i dx^i \, dt
\label{swzmt3}
\end{equation}
where $i=1 \ldots 3$, $\eta_{\mu\nu}$ is the Minkowski metric and
repeated indices are summed over.  To write down the Dirac equation in
this background we start with the standard Dirac $\gamma$-matrices and
from these we define the sets $\{g^\mu(x)\}$ and $\{g_\mu(x)\}$ by
\begin{equation}
g^0 = \gamma^0, \quad g^i = \gamma^i -  \left(
\frac{2GM}{r} \right)^{1/2} \frac{x^i}{r} \gamma^0
\end{equation}
and
\begin{equation} 
g_0 = \gamma^0 + \left( \frac{2GM}{r} \right)^{1/2} \frac{x^i}{r}
\gamma_i , \quad g_i = \gamma_i
\end{equation}
It is straightforward to check that these satisfy
\begin{equation}
\{ g_\mu , g_\nu \} = 2 g_{\mu \nu} I, \quad 
\{ g_\mu , g^\nu \} = 2 \delta_\mu^\nu I
\end{equation}
where $g_{\mu \nu}$ is the metric specified by~\r{swzmt3} and $I$ is
the identity matrix.  In terms of the $g^\mu$ matrices the Dirac
equation is
\begin{equation}
i g^\mu \grad_\mu \psi = m \psi,
\end{equation}
where 
\begin{equation}
\grad_\mu \psi = (\partial_\mu + \frac{i}{2} \Gamma^{\alpha \beta}_{\mu}
\Sigma_{\alpha \beta}) \psi,
\qquad
\Sigma_{\alpha \beta} = \frac{i}{4}[\gamma_\alpha, \gamma_\beta],
\end{equation}
and the components of the spin connection are found in the standard
way (see~\cite{nak-geom}, for example).  With our choice of matrices
we find that the Dirac equation takes the simple form
\begin{equation}
i \!\! \delsl \psi - i \gamma^0 \left(\frac{2GM}{r} \right)^{1/2}
\left( \frac{\partial}{\partial r}  + \frac{3}{4r} \right) \psi = m \psi.
\label{DE}
\end{equation}
Here $\delsl = \gamma^\mu \partial_\mu$ is the familiar Dirac
derivative operator in Minkowski spacetime.  The gravitational effects
are contained in a single interaction term in an analogous manner to
the Coulomb interaction.  The main difference is the presence of a
radial derivative.  This form of the equation is clearly ideal for
scattering calculations as the interaction can be treated
perturbatively.  In the asymptotic region the metric coordinates all
agree with the standard Minkowski interpretation, so there is no
ambiguity in the meaning of any cross sections computed.

\section{Non-Relativistic Approximation}

Before studying the relativistic cross section it is useful to first
consider the non-relativistic limit of the Dirac equation.  The
Hamiltonian form of~\r{DE} contains a single interaction term
\begin{equation}
\hat{H}_I \psi = i \hbar (2GM/r)^{1/2} r^{-3/4} \partial_r (r^{3/4}
\psi), 
\end{equation}
where dimensional constants have been included.  This interaction
Hamiltonian, which incorporates all (general) relativistic effects, is
independent of the particle mass and so embodies the equivalence
principle.  The interaction term is also independent of the speed of
light, so the non-relativistic approximation of the Dirac equation can
proceed in the standard manner~\cite{itz-quant}.  There are no spin
effects to consider, so we arrive at the Schr\"{o}dinger equation
\begin{equation}
  -\frac{\hbar^2}{2m} \bgrad^2 \psi +  i \hbar (2GM/r)^{1/2}
   r^{-3/4} \partial_r (r^{3/4} \psi) = E \psi
\end{equation}
where we have assumed that we have a stationary state of energy $E$.
To simplify this equation we introduce the phase-transformed variable
\begin{equation} 
\Psi=\psi\exp \left(- i(r/a_{G})^{1/2}\right) 
\end{equation}  
where
\begin{equation}
a_{G}=\frac{\hbar ^{2}}{8GMm^{2}}
\end{equation}
and is (eight times) the gravitational equivalent of the Bohr radius.
The new variable $\Psi$ satisfies the simple equation
\begin{equation}
-\frac{\hbar^2}{2m} \bgrad^2 \Psi - \frac{GMm}{r} \Psi = E \Psi.
\end{equation} 
This is precisely the equation we would expect if we used the
Newtonian gravitational potential, and the solutions for $\Psi$ are
Coulomb wave functions.  The standard arguments about the long range
logarithmic phase effects in Coulomb wave functions apply equally to
the $r^{1/2}$ behaviour, so the cross section can be found in the
conventional way~\cite{merz-qm}.  So, in the non-relativistic limit, the
gravitational differential scattering cross section reduces to the
Rutherford formula.  We therefore expect that the full, relativistic
calculation will give a cross section which reduces to the Rutherford
formula for small velocities.

\section{Scattering Cross Section}

The Dirac equation~\r{DE} is well-suited to an iterative solution in
the standard manner.  We seek a solution of
\begin{equation}
[i \!\! \delsl_2 - B(x_2) - m]S_G (x_2,x_1) = \delta^4
(x_2-x_1)
\end{equation}
where
\begin{equation}
B(x) = i \gamma^0 \left(\frac{2GM}{r} \right)^{1/2} \left(
\frac{\partial}{\partial r} + \frac{3}{4r} \right).
\end{equation}
The iterative solution to this equation is
\begin{align}
& S_G(x_f,x_i) = S_F(x_f,x_i) + \int d^4 x_1 \, S_F(x_f,x_1) B(x_1)
S_F(x_1, x_i) \nn \\ 
&+ \iint d^4 x_1 \, d^4 x_2 \, S_F(x_f,x_1)
B(x_1) S_F(x_1, x_2) B(x_2) S_F(x_2, x_i) + \cdots
\label{itsoln}
\end{align}
where $S_F(x_2,x_1)$ is the (position space) Feynman propagator.  The
interaction term $B(x)$ is independent of time so energy is conserved
throughout the interaction.  Converting to momentum space we define the
amplitude
\begin{equation}
\clm =  \bar{u}_s(\bp_f) V  u_r(\bp_i),
\end{equation}
where
\begin{equation}
V = B(\bp_f,\bp_i) + \int \frac{d^3 k}{(2\pi)^3} B(\bp_f,\bk) \frac{\ksl
+m}{k^2-m^2+i\eps} B(\bk,\bp_i) + \cdots .
\end{equation}
Here $B(\bp_f,\bp_i)$ is the spatial Fourier transform of the
interaction term, bold symbols refer to spatial components only, and
for the spinor terms we follow the conventions of Mandl and
Shaw~\cite{mand-qft}.  In terms of $\clm$ the differential cross
section is given by
\begin{equation}
\frac{d\sigma}{d\Omega} = \left(\frac{m}{2\pi} \right)^2 |\clm|^2.
\end{equation}

The Fourier transform of the interaction term is
\begin{equation} 
B(\bp_2,\bp_1) = (2GM)^{1/2} i \gamma^0 \int d^3 x \, e^{-i \bp_2
 \cdot \bx} \frac{1}{r^{1/2}} \left( \frac{\partial}{\partial r} +
 \frac{3}{4r} \right) e^{i \bp_1 \cdot \bx}
\end{equation}
where bold symbols refer to spatial components only.  To evaluate this
we first write
\begin{equation}
B(\bp_2,\bp_1) = (2GM)^{1/2} i \gamma^0 \left( \frac{3}{4} f(\bp_1-\bp_2) +
\left. \frac{\partial f(\lam \bp_1 - \bp_2 )}{\partial \lam}
\right|_{\lam=1} \right) 
\end{equation}
where
\begin{equation} 
f(\bp) = \int d^3 x \frac{e^{i \bp \cdot \bx}}{r^{3/2}} = \left( \frac{2 \pi
}{|\bp|} \right)^{3/2}.
\end{equation}
We therefore find that the momentum space vertex factor is
\begin{equation}
B(\bp_2,\bp_1) = 3 \pi^{3/2}i (GM)^{1/2} \frac{{\bp_2}^2 -
{\bp_1}^2}{|\bp_2-\bp_1|^{7/2}} \, \gamma^0.
\end{equation}
This vertex factor has the unusual feature of vanishing if the ingoing
and outgoing particles are on-shell, since energy is conserved.  It
follows that the lowest order contribution to the scattering cross
section vanishes.  This is reassuring, as the vertex factor goes as
$\sqrt{M}$, and we expect the amplitude to go as $M$ to recover the
Rutherford formula in the low velocity limit.

Working to the lowest non-zero order in $M$ the
transition amplitude becomes
\begin{equation}
\clm = -9 \pi^3 GM \bar{u}_s(\bp_f) \gamma^0 I_1 \gamma^0
u_r(\bp_i)
\end{equation}
where
\begin{equation} 
I_1 =  \int \frac{d^3 k}{(2 \pi)^3} \, \frac{{\bp_f}^2-\bk^2}
{|\bp_f-\bk|^{7/2}} \frac{\ksl + m}{k^2-m^2+i\eps}
\frac{\bk^2-{\bp_i}^2}{|\bk-\bp_i|^{7/2}}.
\end{equation}
Now
\begin{equation}
k^2 - m^2 = E^2 - \bk^2 - m^2 = \bp^2 - \bk^2,
\end{equation}
where $E$ is the particle energy and $\bp^2 = {\bp_i}^2 = {\bp_f}^2$.
The pole in the propagator is therefore cancelled by the vertex
factors, so there is no need for the $i\eps$ prescription.  The
integral we need to evaluate is therefore
\begin{equation}
I_1 = \int \frac{d^3 k}{(2 \pi)^3} \,
\frac{\bk^2 - \bp^2}{|\bp_f-\bk|^{7/2} |\bk-\bp_i|^{7/2}} 
(\ksl + m),
\label{I1}
\end{equation}
which is evaluated in Appendix~\ref{appA}.  The
result is
\begin{equation}
I_1 = \frac{1}{9 \pi^2 \bq^2} \bigl(2m + 3(\!\psl_f + \psl_i) - 4 E
\gamma^0 \bigr)
\label{I1res}
\end{equation}
where $\bq = \bp_f - \bp_i$.  It follows that 
\begin{align}
\clm &= - \frac{\pi G M}{\bq^2} \bar{u}_s(\bp_f) \bigl( 2m - 3(\psl_f +
\psl_i) + 8 E \gamma^0 \bigr) u_r(\bp_i) \nn \\
&= - \frac{4 \pi G M}{\bq^2}  \bar{u}_s(\bp_f) ( 2E \gamma^0 - m)
u_r(\bp_i),
\end{align}
and the differential cross section is given by
\begin{equation}
\frac{d\sigma}{d\Omega} = \frac{(2GMm)^2}{\bq^4} |  \bar{u}_s(\bp_f) (
2E \gamma^0 - m) u_r(\bp_i)|^2.
\end{equation}
So, despite the complexity of going to second order in the iterative
solution, the result is quite straightforward.  Performing the usual
spin sums gives an unpolarised cross section of
\begin{align}
\frac{d\sigma}{d\Omega} &= \frac{(GM)^2}{2 \bq^4} \mbox{Tr} \{
(\psl_f +m) (2E\gamma^0 -m) (\psl_i +m) (2E\gamma^0 -m) \}
\nn \\
&=  \frac{2(GM)^2}{\bq^4}\bigl( m^2(E^2 - \bp_f \dt \bp_i) + (2E^2 -
m^2)^2 + 4E^2 \bp_f \dt \bp_i  \bigr).
\end{align}
If we let $v=|\bp|/E$ denote the particle velocity, and $\theta$ the
scattering angle, we arrive at the simple expression
\begin{equation}
\frac{d\sigma}{d\Omega} = \frac{(GM)^2}{4 v^4 \sin^4(\theta/2)} \bigl(
1 + 2v^2 - 3 v^2 \sin^2(\theta/2) + v^4 - v^4 \sin^2(\theta/2) \bigr).
\end{equation}
The formula has the satisfying property of being independent of the
particle mass, as one would expect from the equivalence principle.  We
delay a fuller discussion of the properties of this result until after
we have established that it is gauge-invariant.

\section{The Kerr-Schild Gauge}

The iterative scheme employed here suggests a generalisation to
alternative field configurations.  In effect, what we have done is
taken the covariant Dirac equation and re-written it in the form
\begin{equation}
(i\!\! \delsl - m) \psi =  i (\!\! \delsl - g^\mu \grad_\mu) \psi
\end{equation}
and we have interpreted the right-hand side as an interaction term
$B(x)$.  This method will clearly provide a sensible iterative scheme
if the right-hand side contains a single factor of some power of $M$.
If this is not the case, there can be no simple correspondence between
the order of the iterative solution, and the order of $M$ in the
amplitude.  The gauge we have exploited to date has the feature that
$B(x)$ goes as $M^{1/2}$.  The obvious question now is whether we can
do better and find a gauge where the interaction goes as $M$.  This
should then avoid having to integrate over intermediate momenta in a
second order diagram.  Such a gauge is provided by introducing the
Eddington-Finkelstein advanced time coordinate, which can be employed
to convert the Schwarzschild metric to Kerr-Schild
form~\cite{kra-exact},
\begin{equation}
ds^2 = \eta_{\mu\nu} dx^\mu dx^\nu - \frac{2GM}{r} l_\mu l_\nu dx^\mu
dx^\nu,
\end{equation}
where
\begin{equation}
l_\mu = (1, x/r, y/r, z/y).
\end{equation}
For our $g^\mu(x)$ matrices we choose
\begin{equation}
g^0 = \gamma^0 + \frac{GM}{r} (\gamma^0 - \gamma_r), \quad
g^i = \gamma^i - \frac{GM}{r} \frac{x^i}{r} (\gamma^0 - \gamma_r),
\end{equation}
where
\begin{equation}
\gamma_r = \frac{x^i}{r} \gamma_i.
\end{equation}
The reciprocal matrices are found to be
\begin{equation}
g_0 = \gamma^0 - \frac{GM}{r} (\gamma^0 - \gamma_r), \quad
g_i = \gamma_i - \frac{GM}{r} \frac{x^i}{r} (\gamma^0 - \gamma_r).
\end{equation}
In this gauge the Dirac equation becomes
\begin{equation}
i \!\! \delsl \psi + \frac{iGM}{r} (\gamma^0 - \gamma_r)  \left(
\frac{\partial}{\partial t} - \frac{\partial}{\partial r} -
\frac{1}{2r}  \right) \psi = m \psi,
\end{equation}
which achieves our goal of constructing an interaction term of order
$M$.

The momentum space representation of the interaction term is now
\begin{equation}
B(\bp_2, \bp_1) = GM \int d^3 x \, e^{-i \bp_2 \cdot \bx} \frac{1}{r} 
\left( \gamma^0 - \frac{x^i}{r} \gamma_i \right)
\left(-E + i \frac{\partial}{\partial r} +
 \frac{i}{2r} \right) e^{i \bp_1 \cdot \bx}.
\end{equation}
The result of this integral is 
\begin{align}
B(\bp_2, \bp_1) =& -\frac{2 \pi GM}{|\bq|^2}(4E \gamma^0 - \psl_1 -
\psl_2) - \frac{4 \pi GM}{|\bq|^4}({\bp_2}^2 - {\bp_1}^2)(\psl_2 -
\psl_1) \nn \\ 
& + \frac{i \pi^2 GM}{|\bq|^3} \Bigl(({\bp_2}^2 -
{\bp_1}^2)\gamma^0  - 2E(\psl_2 - \psl_1) \Bigr). 
\end{align}
This form of interaction is certainly not as elegant as our earlier
gauge choice, but has the advantage that the first-order term in the
iterative solution gives the O$(M)$ contribution to the amplitude.
Since the final two terms in $B(\bp_2, \bp_1)$ vanish on-shell, only
the first term contributes to the amplitude, and we find
\begin{equation}
\clm =  - \frac{4 \pi G M}{\bq^2}  \bar{u}_s(\bp_f) ( 2E \gamma^0 - m)
u_r(\bp_i),
\end{equation}
precisely as obtained earlier.

This calculation confirms that the differential cross section formula,
to order $M^2$, is gauge invariant.  Since a series for the amplitude
in this gauge can only go in orders of $M$, this suggests that there
can be no $M^{3/2}$ contribution in the earlier gauge setup.  That is,
the third order term in the expansion of~\r{itsoln} must vanish.  This
is confirmed by (somewhat tedious) calculations.

\section{Discussion}

Our calculations in two distinct gauges have confirmed that, to lowest
order in the black-hole mass, the unpolarised differential scattering
cross section for a spin-1/2 particle is given by
\begin{equation}
\frac{d\sigma}{d\Omega} = \frac{(GM)^2}{4 v^4 \sin^4(\theta/2)} \bigl(
1 + 2v^2 - 3 v^2 \sin^2(\theta/2) + v^4 - v^4 \sin^2(\theta/2) \bigr).
\end{equation}
As already commented, this formula is independent of the particle mass
and depends only on the incident velocity.  This confirms that the
equivalence principle is directly encoded in the Dirac equation,
though it remains to be shown whether this holds to all orders.  The
formula also makes it clear that the low velocity limit recovers the
Rutherford formula.  The higher order relativistic corrections are not
obvious, but do agree with the small angle formulae obtained by
Collins \textit{et al.}~\cite{col73}, who found that as $\theta
\mapsto 0$ the classical cross section is given by
\begin{equation}
\frac{d\sigma}{d\Omega} =  \frac{4 (GM)^2}{v^4 \theta^4} (1 + 2 v^2 +
v^4).
\end{equation}
Unlike the Collins \textit{et al.} formula (their equation~12),
however, there is no cubic term in $\theta^{-1}$ in the quantum
result, which is an even function of $\theta$.

The massless limit $m \mapsto 0 $ is also well-defined and leads to
the simple formula
\begin{equation}
\frac{d\sigma}{d\Omega} =  \frac{(GM)^2 \cos^2(\theta/2)}{ \sin^4(
\theta/2)} .
\end{equation}
Again, the low-angle limit recovers the classical formula for the
bending of light.  This result also predicts zero amplitude in the
backward direction, $\theta=\pi$.  Null geodesics produce a
significant flux in the backward direction, and the fact that zero is
predicted here is a diffraction effect for neutrinos which goes beyond
the predictions of geometric optics.  A similar prediction of zero
back-scattering for neutrinos was made in~\cite{futt-sct}.  A more
detailed analysis of the cross section in the backward direction also
reveals a large `glory' scattering~\cite{futt-sct,and00}.  In the
geometric optics limit this is attributable to multiple orbits, and in
the quantum description the glory scattering is described by
higher-order terms in $GM$.  To describe these effects in the present
scheme requires extending to higher order in perturbation theory.
This is currently under investigation.

Extending to higher orders also raises the question of the convergence
of the iterative scheme proposed here.  This is not a straightforward
issue to address as there is no dimensionless coupling constant in the
problem.  Also, it is not clear whether higher-order quantum terms
should still be expected to obey the equivalence principle.  One can
easily formulate desirable criteria for convergence, such as $GME<1$
or $GMEv<1$, but these are too restrictive, given that the low angle
formula we arrive at is expected to be valid for all masses and
velocities.  It would appear that the only way to investigate
convergence is to compute the next order terms in the perturbation
series directly.

This work should also have clarified the importance of working
consistently to the correct order in $M$.  This is particularly clear
in the Schwarzschild gauge, where the interaction term contains
factors of $1-(1-2GM/r)^{1/2}$.  An iterative scheme based on this
gauge choice should expand out $B(x)$ as a series in $M$, and then
keep all of the terms up to the desired order.  Such a scheme is
workable, but has the disadvantage of introducing new vertex terms at
each order in the series solution.  It is straightforward to confirm
that such a scheme reproduces our result for the fermion cross
section, to lowest order.

We can now explain the failure of the Born approximation discussed by
Collins \textit{et al.}~\cite{col73} for the scalar case.  These
authors used a similar technique of viewing the difference between the
true and and flat space metrics as an interaction term, and
constructed the amplitude
\begin{equation}
T(\bp_2,\bp_1) = 2GM \int d^3 x \, e^{-i (\bp_2 - \bp_1) \cdot \bx}
\left( \frac{E^2}{r-2GM} + \frac{\bp_2 \dt \bx \, \bp_1 \dt \bx}{r^3}
\right).
\end{equation}
Applying this in the Born approximation as it stands produces an
unphysical answer, which can be traced to the pole at the horizon.
The resolution is that the correct scheme involves a series expansion
of the amplitude as well so that, working to first order in $M$, we
should compute
\begin{equation}
T(\bp_2,\bp_1) = 2GM \int d^3 x \, e^{-i (\bp_2 - \bp_1) \cdot \bx}
\left( \frac{E^2}{r} + \frac{\bp_2 \dt \bx \, \bp_1 \dt
\bx}{r^3} \right).
\end{equation}
Precisely this  integral is obtained in  both of the  the other gauges
discussed in this  paper, up to terms which  vanish on shell.  Working
to first order in $M$ gives rise to the differential cross section
\begin{equation}
\frac{d\sigma}{d\Omega} = \frac{(GM)^2}{4 v^4 \sin^4(\theta/2)}
(1 + v^2)^2 ,
\end{equation}
appropriate for a scalar field.  Again, we see that the equivalence
principle is obeyed, and the various low-angle and low velocity
approximations are retained.

This work suggests a number of generalisations, the most obvious of
which is to more general black-hole configurations.  In this respect a
start for the Kerr case has already been made in~\cite{D00-kerr},
where the Kerr solution was formulated in a gauge with similar
properties to that employed in the first half of this paper.  In
addition, both gauges discussed here look well set up to give a
proper, quantum description of radiation processes as a particle is
accelerated in a gravitational field.  Classical descriptions of such
processes are notoriously tricky and ambiguous.  A further question is
whether the formalism developed here is also appropriate for the
absorption problem~\cite{san77,unr76}.  This involves modifying the
vertex factor in such a way as to explicitly expose the
non-Hermiticity due to the singularity.  We expect to tackle these
issues in future papers.

\appendix

\section{Evaluation of $I_1$}
\label{appA}

To evaluate the integral $I_1$ of equation~\r{I1} we first displace
the origin in $k$-space by $(\bp_f+\bp_i)/2$ to get
\begin{equation}
I_1 =  \int \frac{d^3 k}{(2 \pi)^3} \,
\frac{(\bk+ (\bp_f+\bp_i)/2)^2 - \bp^2}{|\bq/2-\bk|^{7/2} |\bk+\bq/2|^{7/2}} 
\bigl(\ksl + (\psl_f + \psl_i)/2 - E \gamma^0 +m \bigr).
\end{equation}
We now chose coordinate axis with $\bq$ and $\bp_f+\bp_i$ defining the
3 and 1 directions respectively.  These vectors are orthogonal as the
momenta are on-shell.  We next introduce the spheroidal coordinates
\begin{align}
k_1 &= \alpha\sinh\!u \, \sin\! v \, \cos\!\phi \nn \\
k_2 &= \alpha \sinh\!u \, \sin\! v \, \sin\!\phi  \\
k_2 &= \alpha \cosh\!u  \, \cos\! v \nn
\end{align}
where $0 \leq u < \infty$, $0\leq v \leq \pi$, $0 \leq \phi < 2\pi$
and $\alpha=|\bq|/2$.  Exploiting the symmetry in $\bk$ the integral
reduces to two terms:
\begin{align} 
I_1 =& \frac{1}{2\pi^2 \bq^2} \int_0^\infty du \int_0^\pi dv \, \sinh
\! u \, \sin \! v \left\{ \frac{\sinh^2\! u - \sin^2\! v}{(\sinh^2\! u +
\sin^2\! v)^{5/2}} \bigl(\psl_f + \psl_i +2m \bigr) \right. \nn \\
& + \left. \frac{\sinh^2\! u \, \sin^2\! v}{(\sinh^2\! u +
\sin^2 \!v)^{5/2}} (\psl_f + \psl_i -2 E\gamma^0) \right\}.
\end{align}
These integrals are simple to perform and lead to the result of
equation~\r{I1res}.

\end{document}